\documentclass[10pt]{IEEEtran}  
\usepackage{amsmath,amssymb,amsfonts}
\usepackage{graphicx}
\usepackage{cite}
\usepackage{algorithmic}
\usepackage{textcomp}

\usepackage{hyperref}
\usepackage{booktabs}
\usepackage{mathtools}
\usepackage{multirow}
\usepackage{subfig}  
\usepackage{booktabs}
\usepackage[table,dvipsnames]{xcolor} 
\usepackage[linesnumbered,ruled,vlined]{algorithm2e}
\usepackage{comment}
\usepackage{tabularx}
\usepackage{threeparttable}

\usepackage[none]{hyphenat} 
\usepackage{microtype}      
\usepackage[font=scriptsize]{caption}
\SetAlCapFnt{\scriptsize}
\SetAlCapNameFnt{\scriptsize}

\usepackage[none]{hyphenat} 
\usepackage{microtype}      
\def\BibTeX{{\rm B\kern-.05em{\sc i\kern-.025em b}\kern-.08emT\kern-.1667em\lower.7ex\hbox{E}\kern-.125emX}}

\begin{document}

\title{Learning Optimal Crew Dispatch for Grid Restoration Following an Earthquake}

\author{Farshad Amani, \textit{Graduate Student Member, IEEE}, Faezeh Ardali, \textit{Graduate Student Member, IEEE}, Amin Kargarian, \textit{Senior Member, IEEE}

\thanks{This work was supported in part by the National Science Foundation under Grant ECCS-1944752 and in part by the LSU Institute for Energy Innovation and the LSU Provost’s Fund for Innovation in Research.

F. Amani and A. Kargarian are with the Department of Electrical and Computer Engineering, Louisiana State University, Baton Rouge, LA 70803 USA (e-mail: famani1@lsu.edu, fardal1@lsu.edu, kargarian@lsu.edu), and F. Ardali is with the Department of Industrial Engineering (e-mail: fardal1@lsu.edu).
}
}

\maketitle
\begin{abstract}

Post-disaster crew dispatch is a critical but computationally intensive task. Traditional mixed-integer linear programming methods often require minutes to several hours to compute solutions, leading to delays that hinder timely decision-making in highly dynamic restoration environments. To address this challenge, we propose a novel learning-based framework that integrates transformer architectures with deep reinforcement learning (DRL) to deliver near real-time decision support without compromising solution quality. Crew dispatch is formulated as a sequential decision-making problem under uncertainty, where transformers capture high-dimensional system states and temporal dependencies, while DRL enables adaptive and scalable decision-making. Earthquake-induced distribution network damage is first characterized using established seismic standards, followed by a scenario generation and reduction pipeline that aggregates probable outcomes into a single geospatial impact map. Conditioned on this map, the proposed framework generates second-level dispatch strategies, trained offline on simulated and historical events and deployed online for rapid response. In addition to substantial runtime improvements, the proposed method enhances system resilience by enabling faster and more effective recovery and restoration. Case studies, particularly on the 2869-bus European gas and power network, demonstrate that the method substantially accelerates restoration while maintaining high-quality solutions, underscoring its potential for practical deployment in large-scale disaster response.

\end{abstract}

\section*{Nomenclature}
\addcontentsline{toc}{section}{Nomenclature}
\begin{IEEEdescription}[\IEEEusemathlabelsep\IEEEsetlabelwidth{$Q^{\text{g},j}_{\min},\ Q^{\text{g},j}_{\max}$}]

\item[\emph{Sets and Indices:}]
\item[$\mathcal{N}$] Set of all buses (nodes)
\item[$\mathcal{E}$] Set of all distribution lines (branches)
\item[$\mathcal{G}$] Set of distributed generators (DGs)
\item[$\mathcal{F}$] Set of failed components
\item[$\mathcal{R}$] Set of available repair crews
\item[$i,j$] Index for buses
\item[$l$] Index for distribution lines
\item[$t$] Time index
\item[$f_d$] Index for failed component $d$
\item[$r_m$] Index for repair crew $m$

\item[\emph{Power System Parameters:}]
\item[$\rho_l$] Resistance of line $l$
\item[$\chi_l$] Reactance of line $l$
\item[$S^{\max}_l$] Maximum apparent power capacity of line $l$
\item[$M$] A large positive constant
\item[$\phi_j$] Power factor angle at bus $j$
\item[$v_j^{\min}, v_j^{\max}$] Voltage bounds at bus $j$
\item[$P_{j}^{\text{g},\min},\ P_{j}^{\text{g},\max}$] Active power limits of DG at bus $j$
\item[$Q^{\text{g},\min}_{j},\ Q^{\text{g},\max}_{j}$] Reactive power limits of DG at bus $j$
\item[$P^{\text{load}}_{j,t}$] Active load at bus $j$ at time $t$
\item[$Q^{\text{load}}_{j,t}$] Reactive load at bus $j$ at time $t$
\item[\emph{Power Flow Variables:}]
\item[$v_{i,t}$] Voltage magnitude at bus $i$ at time $t$
\item[$p_{l,t}$] Active power flow on line $l$ at time $t$
\item[$q_{l,t}$] Reactive power flow on line $l$ at time $t$
\item[$u_{i,t}$] Binary variable indicating whether bus $i$ is energized at time $t$
\item[$u_{l,t}$] Binary variable indicating whether line $l$ is energized at time $t$
\item[$p^{\text{g}}_{j,t}$] Active power generated by DG at bus $j$ at time $t$
\item[$q^{\text{g}}_{j,t}$] Reactive power generated by DG at bus $j$ at time $t$
\item[$p^{\text{shed}}_{j,t}$] Load shedding (active) at bus $j$ at time $t$
\item[$q^{\text{shed}}_{j,t}$] Load shedding (reactive) at bus $j$ at time $t$

\item[\emph{Repair Crew Dispatch Variables:}]
\item[$\theta_{d,s,c}^\sigma$] Binary variable: crew $c$ travels from $d$ to $s$ in cluster $\sigma$
\item[$y_{d,c}^\sigma$] Binary variable: crew $c$ visits damaged component $d$ in cluster $\sigma$
\item[$s_{d,\sigma}$] Binary variable: component $d$ assigned to depot $\sigma$
\item[$\tau_{d,t}^\sigma$] Binary variable: component $d$ is repaired at time $t$ in cluster $\sigma$
\item[$k_{d,t}$] Binary variable: component $d$ is operational by time $t$
\item[$\alpha_{d,c}$] Arrival time of crew $c$ at component $d$
\item[$\varphi_d$] Node potential variable to eliminate subtours

\item[\emph{Transformer-Based Restoration Parameters:}]
\item[$(x_{dp}, y_{dp})$] Coordinates of each depots
\item[$t_d$] Repair duration of component $f_d$
\item[$CL_d$] Associated unsupplied load $d$
\item[$\tau_{dd'}$] Travel time from previous location to component $f_d$ by crew $r_m$
\item[$\mathcal{S}_s$] Repair sequence assigned to crew $r_m$
\item[$t_d^{\text{outage}}$] Outage duration for component $f_d$
\item[$T$] Overall restoration completion time
\item[$\gamma$] Weighting factor for restoration time vs. ENS

\item[\emph{Seismic Hazard and Fragility Parameters:}]
\item[$MM$] Moment magnitude of earthquake
\item[$R$] Source-to-site distance
\item[$G_a$] Site-specific features (e.g., soil conditions)
\item[$Y$] Ground motion intensity (e.g., PGA)
\item[$\varepsilon$] Residual error in GMPE
\item[$S_d$] Seismic demand at component $d$
\item[$S_{d_{\text{ds}}}$] Median demand to reach damage state $\text{ds}$
\item[$\beta_{\text{ds}}$] Lognormal standard deviation for damage state
\item[$\Phi(\cdot)$] Standard normal cumulative distribution function
$\text{DS}$ given seismic demand $S_d$

\item[\emph{Scenario Generation and Loss Metrics:}]
\item[$N_{\text{sce}}$] Number of Monte Carlo scenarios
\item[$N_{\text{comp}}$] Total number of network components
\item[$L_s$] System loss in scenario $s$
\item[$\delta_{sd}$] Failure indicator for component $d$ in scenario $s$
\item[$F_L(l)$] Empirical distribution of system losses
\item[$L_T$] Loss level corresponding to return period $T$
\item[$w_1, w_2$] Weights for failure count and ENS

\end{IEEEdescription}
\section{Introduction}

\IEEEPARstart{I}{n} recent years, the frequency and severity of high-impact, low-probability events have increased due to global climate change \cite{sadeghian2025simultaneous}. Among these events, floods, hurricanes, and wildfires have become more common, often causing widespread disruption to power systems \cite{amani2024seismic}. Although earthquakes are less frequent, they pose some of the most severe risks to critical infrastructure due to their sudden onset and widespread physical damage \cite{nazemi2019energy}. Historically, earthquakes have led to large-scale blackouts, long recovery periods, and immense socioeconomic losses. For example, the 2011 Tōhoku earthquake in Japan caused unprecedented damage to power infrastructure, demonstrating the vulnerability of modern grids to seismic events \cite{stasinos2022microgrids}.

Various strategies have been developed to improve the resilience of power systems against such disasters. These approaches generally fall into three categories: (i) hardening the physical network before the event \cite{movahednia2023transmission}, (ii) maintaining operation during the disruption \cite{li2021integrating}, and (iii) accelerating restoration afterward \cite{eshkaftaki2024resilience}. Our work focuses on the third category--restoration--critical for minimizing downtime and ensuring a rapid return to normal operations. Deciding which components to restore after a disruption involves combinatorial optimization, where utilities must determine the optimal sequence for repairing components while coordinating crew assignments and routing logistics \cite{zhang2025deep, amani2024seismic}. The growing integration of distributed energy resources (DERs) and advanced control technologies has made power systems increasingly decentralized and interconnected, adding further complexity to the restoration process \cite{poudel2018critical}. These interdependencies and operational constraints, such as crew availability, travel time, and system stability, render restoration planning an NP-hard problem \cite{umunnakwe2021quantitative, wang2022machine}. This underscores the critical need for fast, real-time decision-making tools capable of identifying optimal restoration strategies that minimize downtime and economic losses while rapidly returning the system to service \cite{arjomandi2020modeling}.

Following a high-intensity earthquake, grid performance degrades sharply, leaving many loads unsupplied and causing substantial economic and societal losses \cite{xie2025resilience}. To mitigate such impacts, prior studies have proposed both preventive and adaptive strategies for enhancing distribution system resilience. Preventive measures include seismic retrofitting, undergrounding lines, introducing network redundancy, and adhering to seismic design standards \cite{sadeghian2025simultaneous}. These are complemented by risk-informed planning using fragility curves and hazard maps, alongside proactive maintenance, inventory management, and crew training. Adaptive strategies like network reconfiguration aim to reroute power, isolate damaged components, and restore critical loads in real time \cite{chegnizadeh2025resiliency}. These approaches often incorporate advanced control, optimization, and automation to enable effective response under uncertainty, highlighting the need for integrated planning that bridges long-term investment with rapid response capabilities.

While preventive and adaptive strategies reduce initial damage, effective post-event recovery is essential for timely service restoration. To coordinate recovery efforts, many studies use mathematical optimization—particularly mixed-integer linear programming (MILP)—to model system-wide constraints, resource availability, and temporal dependencies. MILP enables precise scheduling and prioritization but suffers from computational complexity, which limits its practicality for large-scale or time-sensitive scenarios~\cite{umunnakwe2021quantitative}. Since restoration decisions often need to be made in real-time or near-real-time, especially under rapidly evolving disaster conditions, faster solution techniques are critical. To address this, researchers have proposed heuristic and metaheuristic algorithms, such as genetic algorithms (GA), simulated annealing, and particle swarm optimization, which significantly reduce computation time while delivering near-optimal solutions~\cite{younesi2022trends}. Others have explored decomposition methods to manage problem size and adapt decisions as new data becomes available. These advances support more responsive and scalable restoration planning, especially when integrated with DERs, microgrids, and uncertainty-aware models \cite{akter2024review}.

Crew dispatch and service restoration are often modeled as MILP formulations that capture routing, precedence, and network constraints. As feeders and damage sets grow, these models become computationally expensive, often requiring hours to solve even medium-sized cases \cite{chen2018toward}. To improve tractability, many studies apply relaxations or heuristics that retain most of the optimization structure but cut runtime from hours to minutes \cite{tan2019scheduling}. When uncertainty is considered, two-stage stochastic MILPs with progressive hedging decompose scenarios and coordinate them iteratively, enabling solutions for large networks within tens of minutes rather than days \cite{arif2018optimizing}. Robust optimization provides another path, hedging against bounded variability without enumerating scenarios. These models can typically be solved in minutes on medium-scale systems \cite{wu2022robust}. More recent frameworks jointly decide switching/topology, crew routing, and mobile resource allocation. To remain tractable on realistic feeders, they require linearization and decomposition, which reduces runtimes from hours to the order of minutes \cite{lei2019resilient}.

Although optimization and heuristic methods have advanced post-disaster crew dispatch, they remain limited in providing fast, real-time decision support, as traditional MILP-based approaches often require minutes to hours of computation, making them unsuitable for dynamic restoration environments where timely updates are essential \cite{akter2024review, younesi2022trends, lin2019combined}. To address this, we propose a novel learning-based framework that integrates transformer architectures with DRL. Crew dispatch is inherently a sequential decision-making problem under uncertainty, requiring coordination across evolving system states and interdependent repair actions \cite{arif2018optimizing, dolatyabi2025deep}. Transformers efficiently capture long-range dependencies, high-dimensional system states, and temporal patterns, while DRL enables adaptive and scalable decision-making \cite{friedman2023learning, dolatyabi2025graph}. Together, these models achieve near real-time performance without compromising solution quality. The effectiveness of the proposed framework is demonstrated on the 2869-bus European gas and power network, showcasing its potential for scalable and practical deployment under severe disruption scenarios.

In this work, we first characterize earthquake-induced damage at the component level across the distribution network, following established seismic engineering standards. Using these models, we develop a comprehensive scenario-generation and selection pipeline that captures the most probable—and operationally distinct—damage patterns and aggregates them into a single geospatial impact map for operations. Conditioned on this map, we propose a learning-based framework for post-event crew dispatch that delivers second-level decision support. The approach integrates a transformer architecture—capturing high-dimensional system states and temporal dependencies—with DRL to adapt actions under uncertainty and evolving field information. Trained offline on simulated and historical events and deployed online, the framework produces scalable, time-efficient dispatch plans for distribution grids. Case studies demonstrate that this workflow accelerates restoration and measurably enhances distribution-system resilience.
\section{Earthquake Event Assessment}
\label{sec:impact}
Assessing seismic events and their potential impacts on power systems requires a foundational understanding of earthquake characteristics and the methodologies used to evaluate their influence on infrastructure components. This section presents the core principles and modeling tools that underpin earthquake event assessment, focusing on seismic hazard quantification and vulnerability evaluation.
\vspace{-12pt}
\subsection{Earthquake Metrics and Classification}

Unlike gradual or predictable failures, earthquakes can cause simultaneous damage to multiple components across large geographical areas. To characterize the intensity and impact of an earthquake, several seismic metrics are commonly used~\cite{younesi2022trends}. Peak ground acceleration (PGA) reflects the maximum ground acceleration during the event and is widely used in fragility assessments. Peak ground velocity and spectral acceleration capture velocity-sensitive and frequency-sensitive aspects of ground motion, essential for structures exposed to dynamic loading. Peak ground displacement measures the maximum permanent displacement of the ground and is particularly relevant for long-span or flexible components such as overhead transmission lines. Among these, PGA is the most commonly used metric for assessing the impact of seismic events on rigid infrastructure such as substations and distribution poles \cite{amani2024seismic}.
\vspace{-12pt}
\subsection{Seismic Hazard Characterization and Attenuation Relationships}

The seismic intensity experienced by power system components depends not only on the characteristics of the earthquake itself but also on how seismic waves propagate from the source to the point of interest. This propagation is affected by several factors, including earthquake magnitude, distance from the epicenter, fault rupture geometry, and local soil conditions.

To model this spatial attenuation of seismic energy, ground motion prediction equations (GMPEs)---commonly referred to as attenuation relationships---are used. These empirical models estimate the expected ground motion parameters, such as PGA, as a function of source and site parameters. A widely adopted functional form of GMPEs is given by \cite{nazemi2019energy}:

\begin{equation}
    \ln Y = a + F_1(MM) + F_2(R) + F_3(G_a) + \varepsilon
    \label{eq:AR}
\end{equation}

\begin{figure}[]
    \centering
    \includegraphics[width=0.6\columnwidth]{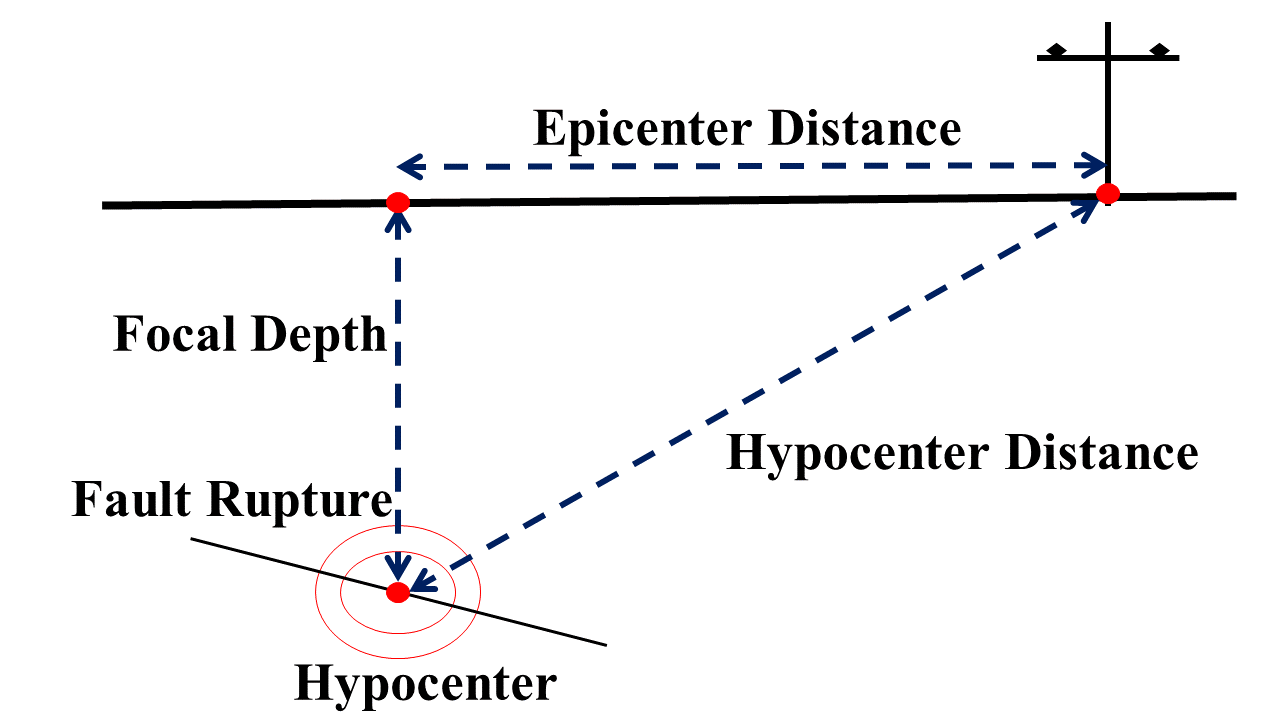}
    \caption{\scriptsize Hypocenter and distance to target component.}
    \label{fig:hypocenter}
\end{figure}

The model parameters are determined based on regional seismic and soil conditions. Using these parameters, the PGA at each component location is calculated according to~\eqref{eq:AR}. Once the seismic intensities are obtained, fragility curves are used to evaluate the probability of component failure. Fig.~\ref{fig:hypocenter} illustrates the earthquake hypocenter and the distance from the focal point to the component of interest.

\vspace{-12pt}
\subsection{Earthquake Scenario Identification}

Earthquake hazard modeling is inherently uncertain due to the stochastic nature of seismic events. Key sources of uncertainty include the exact location of the hypocenter, fault rupture mechanisms, focal depth, and magnitude of the event. Additionally, variations in local soil conditions and ground motion amplification can impact the seismic intensity experienced by infrastructure. Several studies have developed methodologies to incorporate and manage these uncertainties in seismic risk assessment. For instance, \cite{yang2020seismic} and~\cite{nazemi2019seismic} have used probabilistic seismic hazard analysis and the Monte Carlo simulation to generate synthetic earthquake catalogs and simulate the resulting system-wide impact on infrastructures, including power distribution systems.

Since the primary focus of this paper is not seismic hazard characterization, we simplify the modeling by assuming that key seismic event parameters like the earthquake epicenter, magnitude, and fault type are known. To evaluate the performance of the proposed resilience framework, we simulate three distinct earthquake magnitudes representing low, moderate, and high seismic intensities. While this assumption streamlines the analysis, several uncertainties influence the system-level response. One critical task is to estimate the intensity of ground motion at each component's location based on the known event parameters. Once the PGA is calculated at each location, the next step involves translating this ground motion into component-level failure probabilities. Fragility curves are widely used to represent the seismic vulnerability of infrastructure components. These curves provide the conditional probability that a component will reach or exceed a certain damage state given a seismic demand \( S_d \), typically modeled using a lognormal distribution:

\begin{equation}
    FC[\text{ds} \mid S_d] = \Phi\left( \frac{1}{\beta_{\text{ds}}} \ln\left( \frac{S_d}{S_{d_{\text{ds}}}} \right) \right)
    \label{eq:pga}
\end{equation}

These fragility functions have been developed using historical earthquake data, laboratory shake-table tests, and post-disaster field surveys \cite{nazemi2019seismic}. Table~\ref{tableFragility} summarizes the failure probability parameters adopted for various components in the distribution system \cite{amani2024seismic, nazemi2019energy}.

\begin{table}[]
\centering
\caption{\scriptsize Fragility Parameters for Distribution System Components}
\label{tableFragility}
\scriptsize
\begin{tabular}{lccc}
\toprule
\textbf{Parameter} & \textbf{DGs} & \textbf{Substation} & \textbf{Feeders} \\
\midrule
\( S_{d_{\text{ds}}} \) (g) & 0.4 & 0.5 & 0.3 \\
\( \beta_{\text{ds}} \)     & 0.6 & 0.5 & 0.7 \\
\bottomrule
\end{tabular}
\end{table}

Using these parameters, the probability of failure for each component is calculated and used as input to the earthquake failure assessment framework presented in subsequent sections.

\section{Scenario Generation and Selection Strategy}
\label{sec:framework}
We present a structured method for generating and selecting earthquake impact scenarios in power distribution networks, aiming to efficiently capture both typical and extreme outcomes. Using fragility curves and a ground motion model~\eqref{eq:AR}, we define three earthquake intensity levels (magnitudes 6.5, 7.5, and 8.5) and compute the failure probabilities of network components. Monte Carlo simulation is then used to generate \( N_{\text{sce}} \) diverse damage scenarios per intensity level. This process, detailed in Phase~1 of Algorithm~\ref{alg:scenario_generation}, enables realistic scenario modeling without excessive computational burden.
\subsection{Network Performance Evaluation}
After identifying the failed components in each scenario, we perform power flow calculations to assess the system-level impact. The analysis estimates additional consequences, such as energy not supplied (ENS). In each scenario, the failed components are removed from the network model, and a load flow analysis is performed to evaluate resulting outages, islanding, or load shedding. The total system loss can then be characterized as a combination of the number of failed components and the ENS. A weighted formulation is adopted to account for the different natures and units of system impacts. The weighting factors in~\eqref{eq:loss} allow balancing the contribution of structural failures and service interruptions. Hence, this approach provides a meaningful assessment of earthquake consequences.

\label{subsec:powerflow}

\begin{equation}
\text{System Loss} = w_1 \times \text{Failed Components} + w_2 \times \text{ENS (MW)}
\label{eq:loss}
\end{equation}
where \(w_1\) is the weight assigned to the number of failed components and \(w_2\) is the weight assigned to the ENS. \begin{equation}
L_s = \sum_{d=1}^{N_{\text{comp}}} \delta_{sd},
\end{equation}
where \( \delta_{sd} = 1 \) if component \( d \) fails in scenario \( s \), and \( 0 \) otherwise. The empirical distribution function \( F_L(l) \) of system losses is then constructed, allowing identification of loss thresholds corresponding to target return periods. The loss level \( L_T \) associated with a return period \( T \) is determined by:
\begin{equation}
\Pr[L \geq L_T] = \frac{1}{T}.
\end{equation}
For each return period, we select the scenario whose loss \( L_s \) is closest to \( L_T \) and has the highest occurrence probability, ensuring that typical and critical events are well represented.

After securing these representative scenarios, we apply a forward selection approach to reduce the total number of scenarios. At each step, the scenario that most improves the coverage of the original loss distribution is added, minimizing the mismatch between the full and reduced empirical distributions. This approach balances diversity and severity while maintaining computational tractability. Algorithm~\ref{alg:scenario_generation} presents the detailed procedure for generating, reducing, and selecting earthquake impact scenarios.

{\scriptsize
\renewcommand{\thealgocf}{\Roman{algocf}}
\begin{algorithm}[]
\scriptsize  
\caption{Scenario Generation and Reduction Procedure}
\label{alg:scenario_generation}
\KwIn{PGA levels, fragility curves, number of simulations $N_{\text{sim}}$, network data, ground motion model.}
\KwOut{Reduced, weighted set of earthquake impact scenarios.}

\textbf{Phase 1: Damage Scenario Generation}\;
\For{each PGA level $\text{PGA}_k$}{
    \For{$s = 1$ to $N_{\text{sim}}$}{
        \For{each component $d$ in the network}{
            Compute failure probability $FP_d$ using fragility curve at $\text{PGA}_k$\;
            Generate random number $r \sim \mathcal{U}(0,1)$\;
            \eIf{$r < FP_d$}{
                Mark component $d$ as \textbf{failed} in scenario $s$\;
            }{
                Mark component $d$ as \textbf{operational} in scenario $s$\;
            }
        }
        Calculate system loss $L_s$ using~\eqref{eq:loss}\;
    }
}
\vspace{0.3em}
\textbf{Phase 2: Scenario Reduction and Selection}\;
Build empirical CDF of system losses $\{L_s\}$\;
Identify target losses for return periods\;
Select representative scenarios closest to each target loss\;
Apply forward scenario selection to reduce redundancy in the scenario set\;
\textbf{Return} reduced and weighted scenario set\;
\end{algorithm}
}
\vspace{-12pt}
\vspace{-10pt}
\subsection{Power Flow Calculation}
Several important system-level conditions must be considered to calculate the ENS following an earthquake. Earthquakes can cause widespread disruptions in the distribution network, including islanding, the formation of multiple electrically isolated subnetworks, and partial disconnection of communities from the main grid. Additional challenges may include the inaccessibility of certain substations and the failure or shutdown of distributed generator (DG) units due to safety protocols or physical damage.

To model these contingencies, a power flow formulation is adopted that incorporates the possibility of component failures, such as lines, DGs, and substations, using binary decision variables. This enables the evaluation of network survivability and operational feasibility under various post-disaster scenarios, forming the basis for reliable ENS assessment.

The model includes key operational constraints such as nodal power balance, branch flow limits, prioritized load shedding, voltage magnitude bounds, and DG output limits based on unit availability.
\text{1) Nodal Power Balance (for all $j \in \mathcal{N}$, $t \in T$):}
\begin{align}
p^{\text{g}}_{j,t} - \left(P^{\text{load}}_{j,t} - p^{\text{shed}}_{j,t}\right) 
&= \sum_{k \in \Omega^{\text{in}}(j)} p^{f}_{k,t} - \sum_{m \in \Omega^{\text{out}}(j)} p^{f}_{m,t}, \label{eq:active_balance} \\
q^{\text{g}}_{j,t} - \left(Q^{\text{load}}_{j,t} - q^{\text{shed}}_{j,t}\right) 
&= \sum_{k \in \Omega^{\text{in}}(j)} q^{f}_{k,t} - \sum_{m \in \Omega^{\text{out}}(j)} q^{f}_{m,t}. \label{eq:reactive_balance}
\end{align}

\text{2) Branch Flow Constraints (for all $l \in \mathcal{E}$, $t \in T$, $i,j \in \mathcal{N}$):}

\begin{align} - M \left(1 - u_{l,t} \cdot u_{j,t} \cdot u_{i,t} \right) &\leq v_{i,t} - v_{j,t} - \left(\rho_l p_{l,t} + \chi_l q_{l,t} \right) \notag \\ &\leq M \left(1 - u_{l,t} \cdot u_{j,t} \cdot u_{i,t} \right), \label{eq:branch_voltage_diff} \\[4pt] - S_l^{\max} \cdot \left(u_{l,t} \cdot u_{j,t} \cdot u_{i,t}\right) &\leq p_{l,t} \leq S_l^{\max} \cdot \left(u_{l,t} \cdot u_{j,t} \cdot u_{i,t}\right), \label{eq:branch_active_limit} \\[4pt] - S_l^{\max} \cdot \left(u_{l,t} \cdot u_{j,t} \cdot u_{i,t}\right) &\leq q_{l,t} \leq S_l^{\max} \cdot \left(u_{l,t} \cdot u_{j,t} \cdot u_{i,t}\right). \label{eq:branch_reactive_limit} \end{align}

\text{3) Load Shedding Constraints:}
\begin{align}
0 \leq p^{\text{shed}}_{j,t} &\leq P^{\text{load}}_{j,t}, && \forall j \in \mathcal{N},\ \forall t \\
q^{\text{shed}}_{j,t} &= p^{\text{shed}}_{j,t} \cdot \tan(\phi_j), && \forall j \in \mathcal{N},\ \forall t 
\label{binary}
\end{align}

\text{4) Voltage Magnitude Constraints:}
\begin{align}
v_j^{\min} \leq v_{j,t} \leq v_j^{\max}, && \forall j \in \mathcal{N},\ \forall t 
\end{align}

\text{5) DG Output Constraints:}
\begin{align}
P^{\text{g},\min}_{j} \cdot u_{j,t} \leq p^{\text{g}}_{j,t} &\leq P^{\text{g},\max}_{j} \cdot u_{j,t}, && \forall j \in \mathcal{G},\ \forall t  \\
Q^{\text{g},\min}_{j} \cdot u_{j,t} \leq q^{\text{g}}_{j,t} &\leq Q^{\text{g},\max}_{j} \cdot u_{j,t}, && \forall j \in \mathcal{G},\ \forall t 
\end{align}

The power flow model described above forms the operational backbone for assessing ENS. Once damaged components are identified, repair crew dispatch decisions—whether determined classically or via machine learning—directly impact network recovery and load restoration timing.

\subsection{Model Linearization}
Branch flow constraints \eqref{eq:branch_voltage_diff}, \eqref{eq:branch_active_limit}, and \eqref{eq:branch_reactive_limit} contain nonlinear product terms. To linearize the product of binary variables, we introduce auxiliary binary variables $z \in \{0,1\}$ that represent these products. Consider the nonlinear term $u_{l,t} \cdot u_{i,t} \cdot u_{j,t}$, which involves the product of three binary variables. We linearize such expressions in a pairwise manner. Let $z_2 = u_{l,t} \cdot u_{i,t}$, so that $z_1 = z_2 \cdot u_{j,t}$, where $z_1$ represents the original three-variable product. Both $z_1$ and $z_2$ can then be linearized as follows:
\begin{align}
z_1 &\leq u_{l,t}; z_1 \leq u_{i,t}; z_1 \geq u_{l,t} + u_{i,t} - 1 \label{eq:lin3}\\
z_2 &\leq z_1; z_2 \leq u_{j,t}; z_2 \geq z_1 + u_{j,t} - 1 \label{eq:lin6}\\
z_1 &, z_2 \in \{0,1\} \label{eq:lin7}
\end{align}
\subsection{Repair Crew Dispatch Optimization}

The repair crew dispatch problem is modeled as a vehicle routing problem, where depots coordinate the sequential dispatch of multiple crews to repair damaged components. The process begins with a clustering step, where each damaged component \( d \in DN \) is assigned to the nearest depot \( \sigma \in DP \) by minimizing the total distance:

\begin{align}
\min \sum_{d \in DN} \sum_{\sigma \in DP} \text{distance}(d, \sigma) \cdot s_{d,\sigma} \label{eq:clustering_obj} \\
\text{s.t.} \quad \sum_{\sigma \in DP} s_{d,\sigma} = 1, \quad \forall d \in DN \label{eq:clustering_assign}
\end{align}

Here, \( s_{d,\sigma} \in \{0,1\} \) indicates whether component \( d \) is clustered to depot \( \sigma \). Each depot then dispatches a set of repair crews \( c \in C_\sigma \) to visit and repair the assigned components. The dispatch is modeled using binary decision variables:
\begin{itemize}
    \item \( \theta_{d,s,c}^\sigma = 1 \) if crew \( c \) travels from node \( d \) to \( s \)
    \item \( y_{d,c}^\sigma = 1 \) if crew \( c \) visits node \( d \)
\end{itemize}

The routing constraints include:
\begin{align}
\sum_{c \in C_\sigma} y_{d,c}^\sigma = 1, \quad \forall d \in DN_\sigma \label{eq:visit_once} \\
\sum_{s \in DN_\sigma \setminus \{d\}} \theta_{d,s,c}^\sigma - \sum_{s \in DN_\sigma \setminus \{d\}} x_{s,d,c}^\sigma = 0, \quad \forall d, c \label{eq:flow_cons}
\end{align}

To prevent subtours, node potential variables \( \varphi_d \in \mathbb{R} \) are introduced:
\begin{align}
\varphi_d - \varphi_s + |DN_\sigma| \cdot \theta_{d,s,c}^\sigma \le |DN_\sigma| - 1, \quad \forall d \ne s, \; c \in C_\sigma \label{eq:subtour}
\end{align}

Arrival times at each node are tracked using continuous variables \( \alpha_{d,c} \), and computed recursively:
\begin{align}
AT_{s,c} \ge \alpha_{d,c} + rt_{d,c} + tt_{d,s,c} - M(1 - \theta_{d,s,c}^\sigma) \label{eq:arrival_time}
\end{align}

The repair time step is modeled using binary variable \( \tau_{d,t}^\sigma \in \{0,1\} \), where:
\begin{align}
\sum_{t \in T} \tau_{d,t}^\sigma = 1, \quad \forall d \in DN_\sigma \label{eq:repair_one_timestep}
\end{align}

The operational status of component \( d \) at time \( t \), denoted by \( k_{d,t} \), is updated based on repair progress:
\begin{align}
k_{d,t} = \sum_{k=1}^{t} \tau_{d,k}^\sigma, \quad \forall d \in DN_\sigma, \; t \in T \label{eq:status_update}
\end{align}


The power system is divided into multiple zones, each assigned to a specific depot. In each zone, repair crews are dispatched from the corresponding depot to service all damaged components marked by red stars. Once the repairs are completed, the crews return to their respective depots, completing the restoration route.


\section{Learning-based Restoration Methodology}

\begin{figure}[]
    \centering
    \includegraphics[width=0.8\columnwidth]{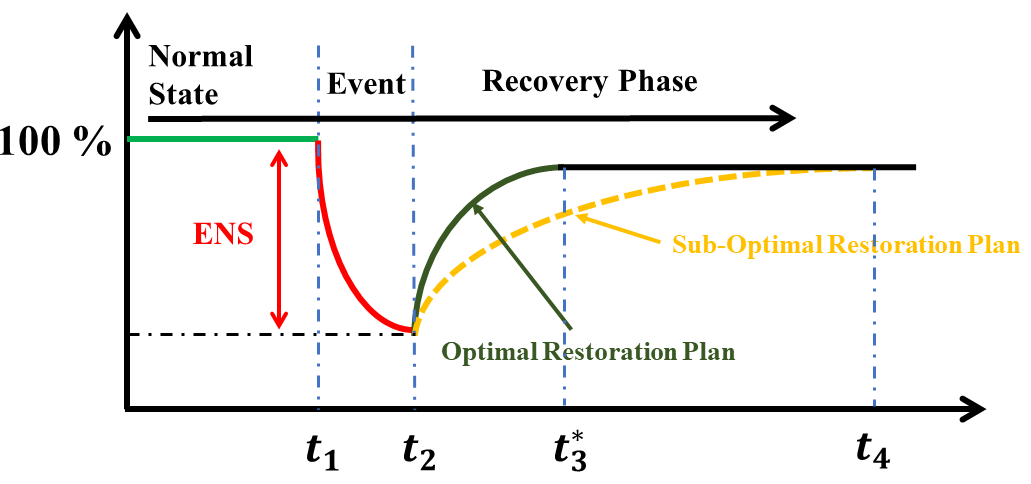}
    \caption{\scriptsize System performance following an earthquake.}
    \label{fig:rescurve}
\end{figure}

Following an earthquake, a significant portion of the system load may become unsupplied due to component failures. As illustrated in Fig.~\ref{fig:rescurve}, system performance experiences an immediate drop at time \( t_1 \) (the onset of disruption). Since earthquakes are short-duration events compared to slower-onset hazards such as floods or hurricanes, restoration activities can begin soon after the disruption at time \(t_2\). If an optimal post-event recovery plan is developed, a faster restoration trajectory can be achieved, reaching a near-normal performance level at time \( t_3^* \). Without such planning, the recovery process is slower, and complete system restoration is delayed.

In post-earthquake restoration, repair crews are dispatched from multiple depots to service failed components, with each crew operating sequentially and returning only after completing its tasks. Varying crew availability, spatial failure distribution, travel times, and repair durations make this a large-scale, NP-hard combinatorial problem. To address this, we propose a transformer-based sequential decision-making framework integrated with DRL. Leveraging the transformer's ability to model variable-length sequences and dependencies via self-attention \cite{friedman2023learning}, the model learns efficient repair sequences that minimize downtime and adapt to new scenarios—unlike classical MILP approaches that rely on explicit optimization.

\subsection{Problem Formulation}
Algorithm~\ref{alg:scenario_generation} provides a set of damage scenarios, each representing a post-earthquake state of the distribution network characterized by a set of failed components $\mathcal{F_l} = \{f_1, f_2, \dots, f_n\}$. Each failed component $f_d$ is located at coordinates $(x, y)$, associated with a repair duration $t_d$, and an unsupplied load value $CL_d$. A set of available repair crews is represented by $\mathcal{R} = \{r_1, r_2, \dots, r_m\}$, where each crew $r_m$ is initially stationed at a dispatch location $(x_{dp}, y_{dp})$.

The objective is to determine the optimal repair sequences $\mathcal{S}_s$ for each crew $r_m$, aiming to minimize the overall restoration time and the total ENS during the recovery process. The restoration completion time $T$ is defined as:

\begin{equation}
T = \max_{s \in \mathcal{R}} \left\{ \sum_{f_d \in \mathcal{S}_s} \left( \tau_{{f_d}{f_d'}} + t_{f_d} \right) \right\},
\label{eq:restoration_time}
\end{equation}

This ensures that the objective reflects not just individual crew efficiency, but also overall system recovery completeness. $\tau_{{f_d}{f_d'}}$ denotes the travel time from the previous location (either the dispatch depot or the last repaired component) to the failed component $f_d$. 

In this formulation, the travel and repair times are sequentially accumulated along each crew's repair path $\mathcal{S}_s$, and the system restoration time $T$ is determined by the longest path among all crews. The ENS associated with each component is accumulated over time until the component is repaired. To balance these two objectives, the restoration problem is formulated as a weighted multi-objective optimization:

\begin{equation}
\min \left( \gamma \, T + (1-\gamma) \sum_{o=1}^{n} CL_d \, t_d^{\text{outage}} \right),
\label{eq:multiobjective}
\end{equation}

Here, \( t_s^{\text{outage}} \) denotes the duration between the time of the earthquake and the completion of repairs for component \( f_d \). During this outage window, the associated load \( CL_d \) remains unsupplied, directly contributing to the cumulative ENS. $\gamma \in [0,1]$ is a weighting factor that controls the trade-off between minimizing restoration time and minimizing total ENS. The objective function, therefore, balances timely restoration with minimizing prolonged service disruption, enabling prioritization of critical components.

\subsection{Transformer-Based Methodology}

We use a transformer encoder-decoder model due to its strong ability to capture sequential dependencies. Unlike traditional recurrent neural networks, transformers do not rely on step-by-step processing of input sequences. Instead, they use a fully parallelizable structure that leverages self-attention mechanisms to model relationships between elements in a sequence, regardless of their distance. The architecture consists of two main components: an encoder, which transforms the input into a contextual representation, and a decoder, which uses this representation to generate the output sequence. The input to the encoder includes the initial conditions represented by a vector $\psi_{f_d}$ for each failed component:
\begin{equation}
\psi_{f_d} = (x_{dp}, y_{dp}, t_{f_d}).
\end{equation}

The encoder extracts internal features from the input using a multi-head self-attention mechanism, which assigns weights to different elements, and a position-wise feed-forward network, which applies non-linear transformations independently to each position. This generates context vectors embedding the spatial and temporal characteristics of failed components. The decoder, with a similar structure, includes an additional attention mechanism to reference encoder outputs and iteratively construct optimal repair sequences. At each step, it outputs probabilities for assigning crews to components based on the current restoration state, while masking already scheduled crews. This architecture captures long-range dependencies more effectively than traditional sequence models.

\begin{figure}[h]
    \centering
    \includegraphics[width=1\columnwidth]{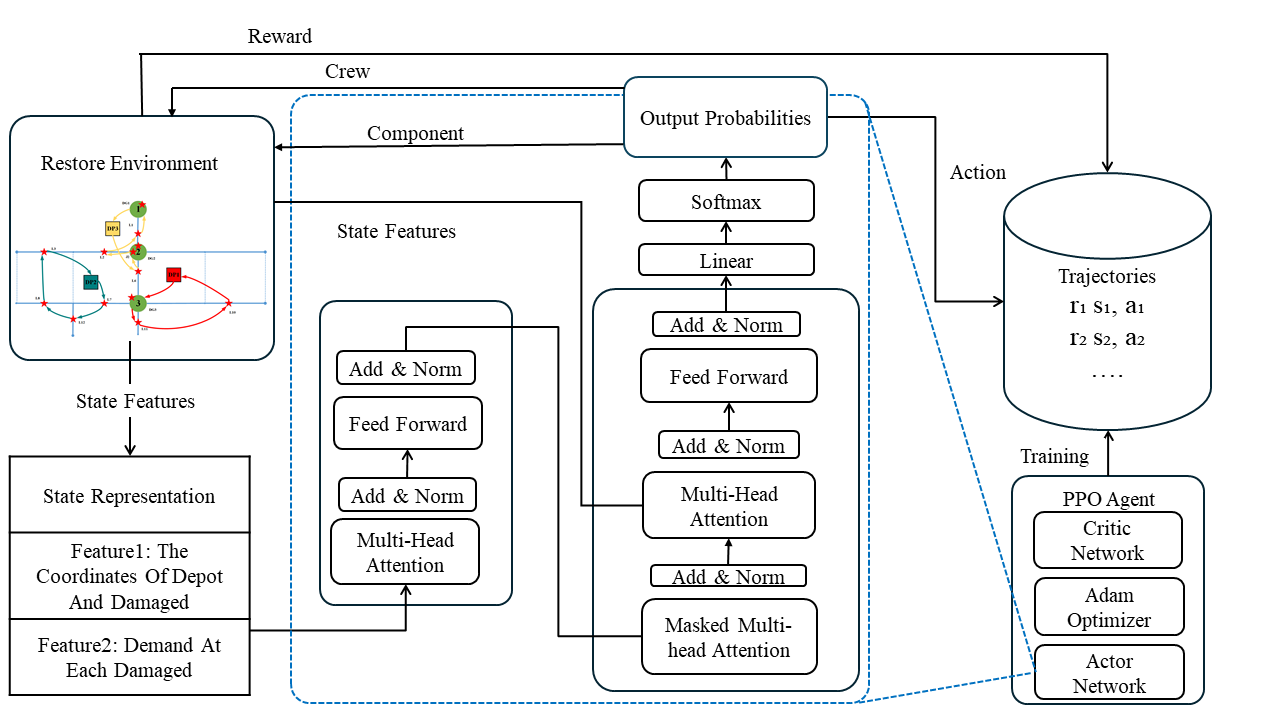}
    \caption{\scriptsize Proposed DRL-transformer framework for post-disaster repair scheduling.}
    \label{fig:Transformer}
\end{figure}

To further optimize sequences through experience, the Transformer is integrated with DRL. Using proximal policy optimization (PPO), the agent interacts with the environment, storing state–action–reward tuples in a replay buffer until all jobs are processed. PPO then updates the policy (actor) and value (critic) networks, striking a balance between exploration and exploitation to ensure stable training. This process repeats until convergence, yielding the final DRL scheduling model. Fig.~\ref{fig:Transformer} illustrates the proposed architecture, where the Transformer and reinforcement learning agent jointly generate sequential repair actions.

At the core of the transformer is the attention mechanism, which enables the model to dynamically focus on the most relevant parts of the input. Attention operates on queries (Q), keys (K), and values (V), which are projected from the input and used to compute relevance scores. These scores are used to calculate a weighted sum of the values:

\begin{equation}
\mathrm{Attention}\bigl(Q, K, V\bigr) = \mathrm{softmax}\left( \frac{Q K^\top}{\sqrt{d_k}} \right) V.
\end{equation}
where \(d_k\) is the dimension of the key vectors, and the softmax function ensures the weights sum to one. This mechanism allows the model to prioritize different parts of the sequence based on learned relevance.
To further enhance its representational power, the transformer applies this attention mechanism in parallel across multiple heads, known as multi-head attention. Each head operates on differently projected versions of Q, K, and V, enabling the model to capture various types of relationships simultaneously. The outputs from all attention heads are then concatenated and passed through a final linear transformation:

\begin{equation}
\mathrm{MHA}\bigl(Q,\, K,\, V\bigr) = \mathrm{Concat}\bigl(\mathrm{head}_1, \dots, \mathrm{head}_h\bigr)\, W^{O}.
\end{equation}
where each head is computed as:

\begin{equation}
\mathrm{head}_h = \mathrm{Attention}\bigl(Q \times W_h^{Q},\, K \times W_h^{K},\, V \times W_h^{V}\bigr).
\end{equation}
where \(W_h^{Q}\), \(W_h^{K}\), and \(W_h^{V}\) are learnable projection matrices for the queries, keys, and values in the \(h\)-th head, respectively. \(W^{O}\) is a learnable output projection matrix applied after concatenating the heads. This multi-head attention structure provides richer context modeling and supports learning diverse patterns within the input data. The combination of attention-based processing and parallelization makes the transformer highly effective for tasks involving sequential or structured information.
\subsection{Practical Benefits and Computational Efficiency}

The proposed transformer model is computationally efficient, offering near-real-time solutions for rapid decision-making following earthquakes. The trained model generalizes across diverse scenarios without the need for repeated retraining, thus providing a robust framework for emergency response and recovery planning. By integrating data-driven repair sequencing with physical power flow constraints, the framework balances solution quality with computational speed, making it suitable for planning and real-time operations.

\section{Numerical results}
\label{sec:results}
We have tested the proposed approach on multiple systems, including a real-world large system. A small 13-bus system was first analyzed, followed by larger 123- and 2869-bus systems to assess scalability. Earthquake scenario identification and learning crew dispatch were conducted in Python. Classical crew dispatch optimization and power flow calculations were performed in GAMS using the Gurobi MILP solver. MATLAB was used to implement heuristic approaches.
\subsection{13-bus System}
\begin{figure}[]
    \centering
    \includegraphics[width=0.65\columnwidth]{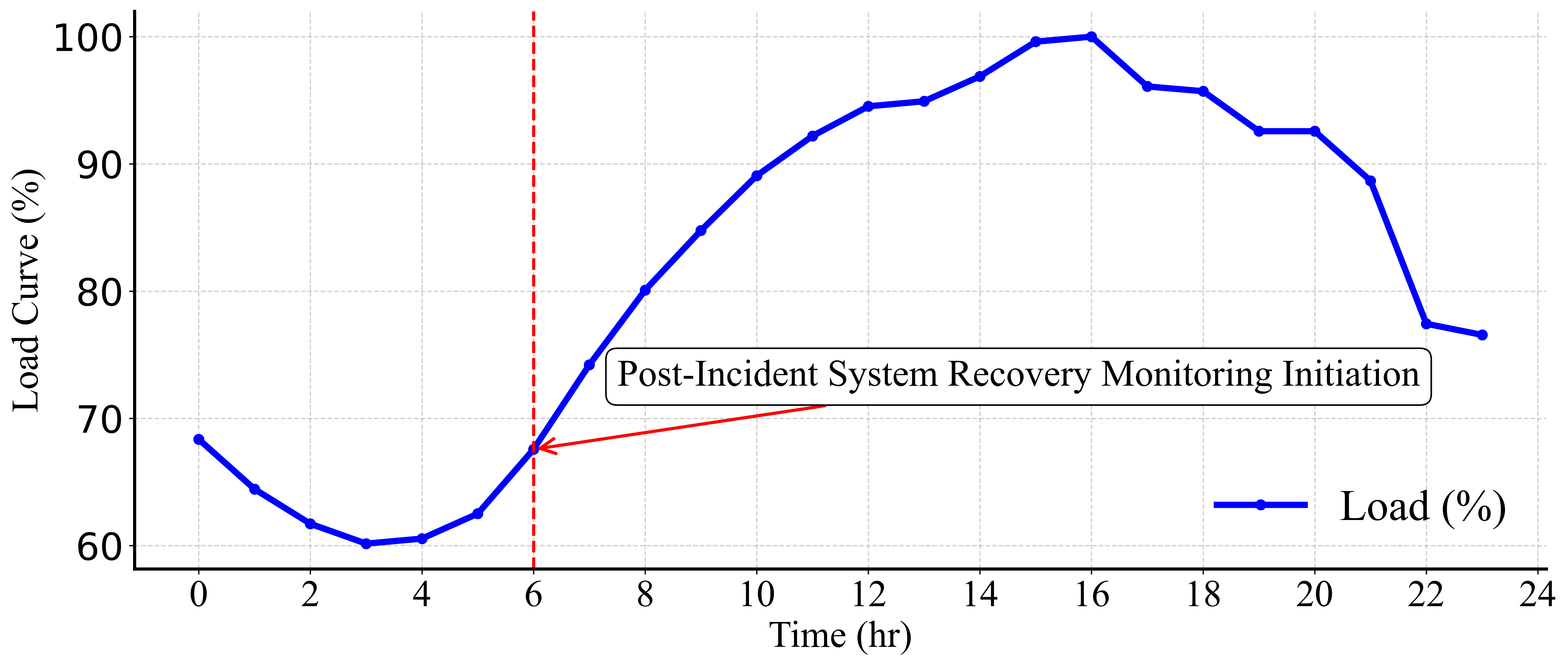}
    \caption{\scriptsize 13 and 123-bus system load curve.}
    \label{fig:load}
\end{figure}

\begin{figure}[]
    \centering
    \includegraphics[width=0.65\columnwidth]{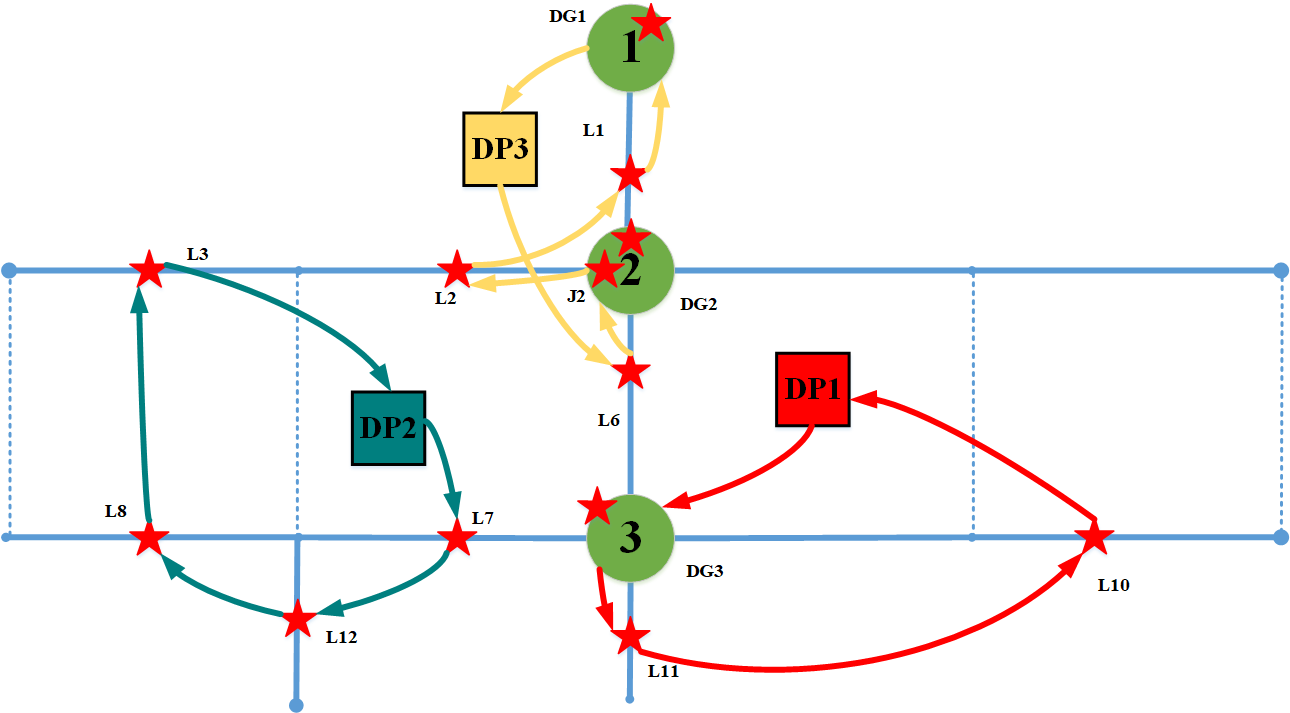}
    \caption{\scriptsize Repair route for 13-bus system restoration.}
    \label{fig:13node}
\end{figure}

Algorithm~\ref{alg:scenario_generation} is first applied to realize the earthquake-induced damage scenarios under three different earthquake magnitudes. We then execute procedures shown in Fig.~\ref{fig:Transformer} to discover the optimal decision to restore the damaged elements to their normal operating state with the least possible total loss. The repair times are assumed to be one hour for distribution lines and two hours for DGs and substations. The baseline load profile is illustrated in Fig.\ref{fig:load}. In Fig.\ref{fig:13node}, the system components are labeled as L (lines), DGs, and J (substations), and the crew dispatch routes are shown under the highest earthquake intensity scenario or 8.5 Richter. A comparative summary of the restoration outcomes across the three intensity levels is presented in Table~\ref{table:13node}.


\begin{table}[h]
\centering
\scriptsize
\caption{\scriptsize Restoration Results of 13-Bus System Under Different Earthquake Scenarios}
\label{table:13node}
\begin{tabular}{lccccc}
\toprule
\textbf{Scenario} & \textbf{Intensity} & \textbf{Load Shed} & \textbf{Objective Value} & \textbf{Comp. Time} \\
         & \textbf{(R)}       & \textbf{(MWh)}     &            & \textbf{(sec)}         \\
\midrule
Sce. 1-1   & 6.5       & 6.3       & 16049        & 11          \\
Sce. 1-2   & 7.5       & 18.88      & 48095    &  12          \\
Sce. 1-3   & 8.5       & 36.46      & 92879         & 12          \\
\bottomrule
\end{tabular}
\end{table}

\begin{figure}[]
    \centering
    \includegraphics[width=0.65\columnwidth]{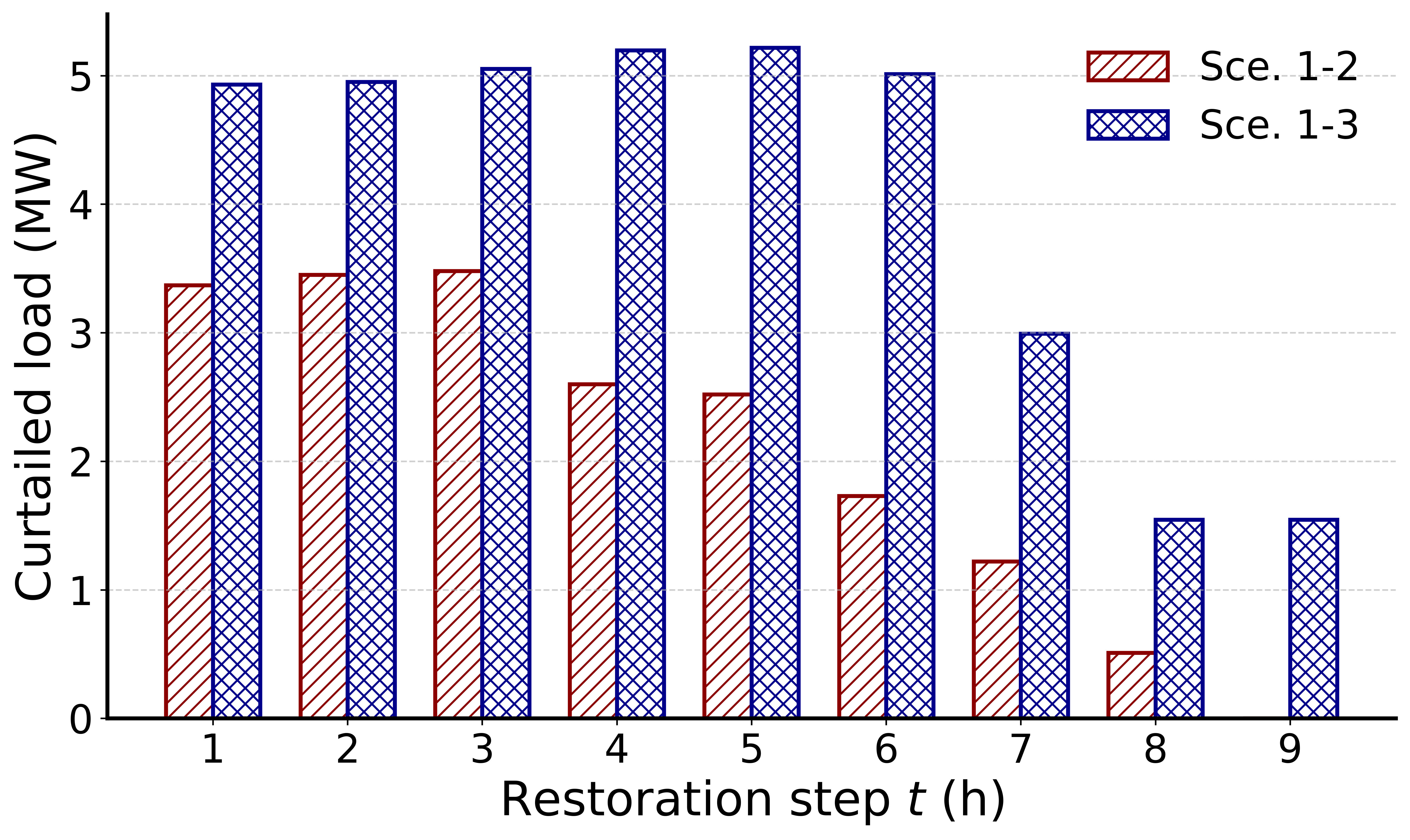} 
    \caption{\scriptsize Curtailed load for Scenarios 2 and 3 in the 13-bus  system.}
    \label{fig:13recovery}
\end{figure}

The curtailed load during the recovery process for scenarios 1-2 and 1-3 is illustrated in Fig. \ref{fig:13recovery}. The curves initially exhibit an ascending trend due to the increasing demand following an earthquake, aligned with the daily load profile. In the early hours of restoration, more load is interrupted, resulting in higher ENS. As damaged components are progressively repaired, the system regains its capacity to serve the load, leading to a decline in curtailed energy. The sequence of component repairs at each time step is detailed in Table~\ref{table:repair_sequence}. 
\begin{table}[]
\centering
\caption{\scriptsize Repaired Components at Each Time Step During Restoration}
\label{table:repair_sequence}
\scriptsize
\resizebox{\columnwidth}{!}{%
\begin{tabular}{lcccccccccc}
\toprule
\textbf{Time} & T1 & T2 & T3 & T4 & T5 & T6 & T7 & T8 & T9 & T10 \\
\midrule
\textbf{Element} 
& L3, L5 
& DG1 
& z2, z3 
& L2 
& L8, L12 
& z1 
& J2, L1, L6 
& DG2, DG3, z5 
& L7 
& L11 \\
\bottomrule
\end{tabular}%
}
\end{table}

To highlight the effectiveness of the proposed transformer-based method, we compare its runtime performance against two commercial solvers, CPLEX and Gurobi, as well as a GA. Based on the objective values obtained by each method, we observe that all solutions are close to the optimal value. Notably, the learning crew dispatch achieves a near-optimal solution with an error of less than 2.1\%. The proposed learning-based crew dispatch achieves better results in the last two scenarios. The detailed results are presented in Table~\ref{table:obj}.

\begin{table}[h]
\centering
\scriptsize
\setlength{\tabcolsep}{16pt}
\caption{13-Bus System Objective Value Error of Learning-Based Crew Dispatch Compared to CPLEX, Gurobi, and GA}
\label{table:obj}
\begin{tabular}{lccc}
\toprule
\textbf{Scenario} & \textbf{CPLEX} & \textbf{Gurobi} & \textbf{GA} \\
\midrule
Sce. 1-1 & 0.3\% & 0.3\% & 2.1\% \\
Sce. 1-2 & 1.1\% & 1.1\% & -2.3\% \\
Sce. 1-3 & 1.4\% & 1.4\% & -1.8\% \\
\bottomrule
\end{tabular}
\end{table}
\vspace{-10pt}
\subsection{123-bus System}
We introduced additional complexity for the 123-bus test system compared to the smaller 13-bus system by allowing each depot to host a different number of repair crews. Five depots were considered, with DP1 having three crews, DP2 and DP3 each having two crews, DP4 having one crew, and DP5 having two crews. Despite the increased scale and complexity, the learning crew dispatch approach successfully generated efficient repair schedules. We applied Algorithm~\ref{alg:scenario_generation} to generate the earthquake-induced damage scenarios and perform system restoration. The same naming conventions and repair time assumptions were used for consistency. A summary of the restoration outcomes for this system is provided in Table~\ref{table:123node}.

\begin{table}[h]
\centering
\scriptsize
\renewcommand{\arraystretch}{1.1}
\setlength{\tabcolsep}{3pt} 
\caption{Restoration Results Under Different Earthquake Scenarios for 123-Bus System}
\label{table:123node}
\begin{tabular}{lccccc}
\toprule
\textbf{Scenario} & \textbf{Intensity (R)} & \textbf{Load Shed (MWh)} & \textbf{Objective Value} & \textbf{Comp. Time (sec)} \\
\midrule
Sce. 2-1 & 6.5 & 12.30  & 22,879  & 13 \\
Sce. 2-2 & 7.5 & 30.89  & 57,457  & 13 \\
Sce. 2-3 & 8.5 & 40.90  & 76,076  & 14 \\
\bottomrule
\end{tabular}
\end{table}

\begin{figure}[]
    \centering
    \includegraphics[width=0.65\columnwidth]{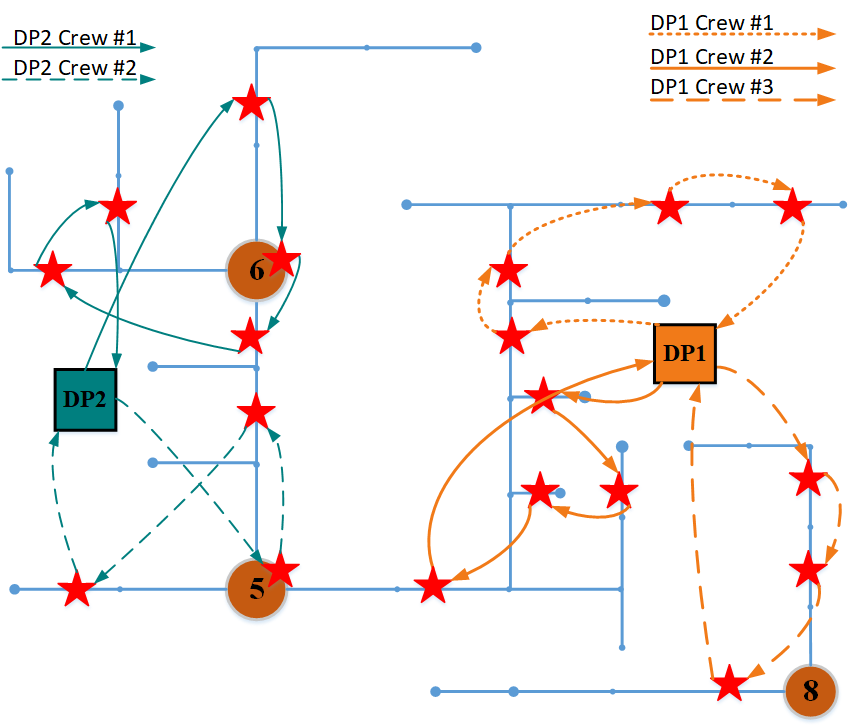}
    \caption{\scriptsize Repair route for 123-bus system restoration.}
    \label{fig:123node}
\end{figure}

Given the large size of the system and the high number of damaged components, we illustrate only a portion of the restoration process for clarity. We focus on the area where depots DP1 and DP2 dispatch their crews to perform repairs in scenario 2-1. The corresponding crew routes and restoration sequence for this section of the system are presented in Fig.~\ref{fig:123node}. The learning crew dispatch approach demonstrated full compliance with the operational constraints. Each depot exclusively addressed its assigned set of damaged components, ensuring no overlap or violation of the clustering assignment. Furthermore, all dispatched crews followed a feasible route, completing the repair tasks within their designated zones and returning to their respective depots upon completion. This outcome confirms the model’s ability to enforce spatial repair assignments and route closure constraints, supporting the practicality and scalability of the proposed method.

\begin{figure}[]
    \centering
    \includegraphics[width=0.65\columnwidth]{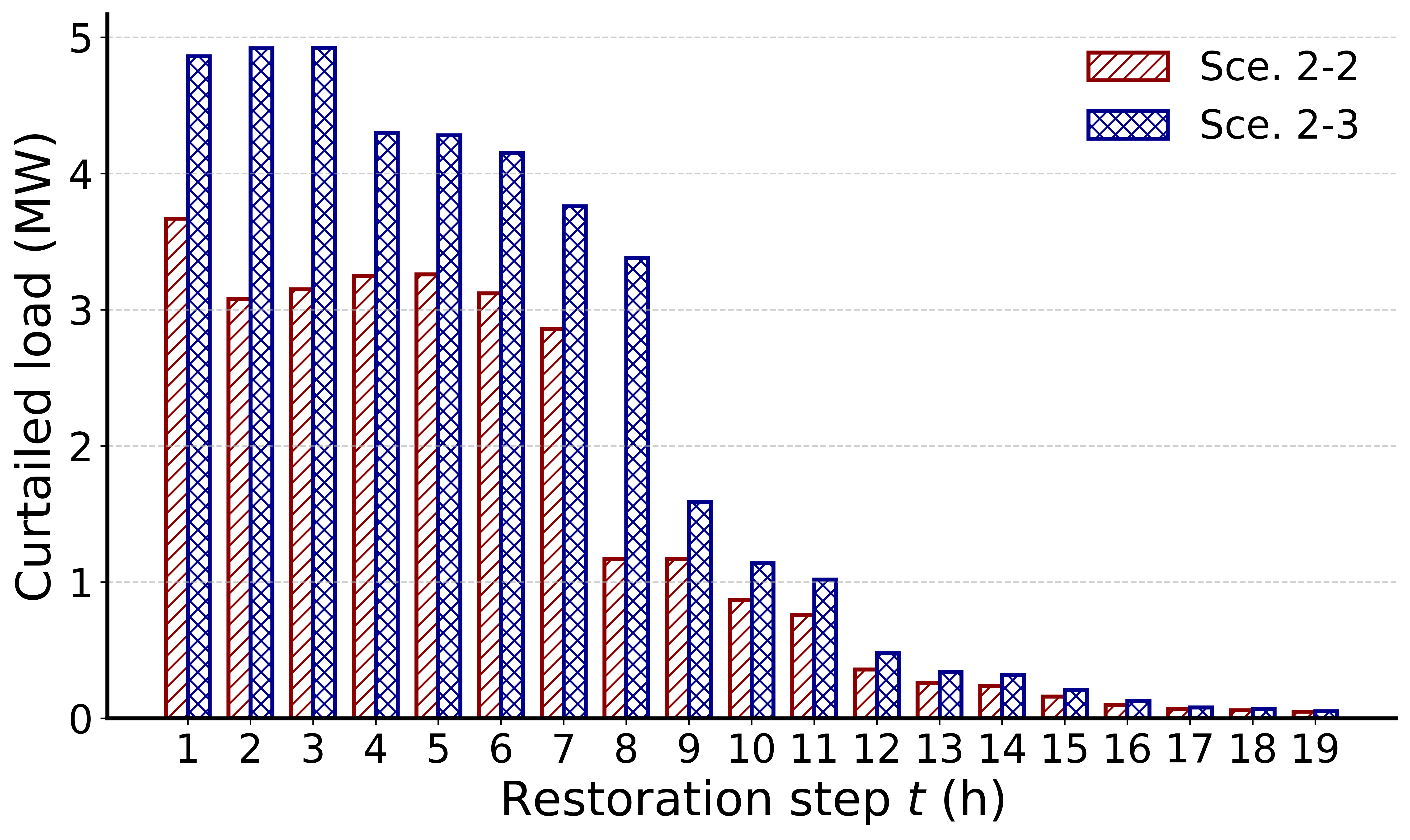} 
    \caption{\scriptsize Curtailed load for Scenarios 2 and 3 in the 123-bus  system.}
    \label{fig:123recovery}
\end{figure}

Fig.~\ref{fig:123recovery} also presents the curtailed load during the recovery process for Scenarios 2-2 and 2-3, demonstrating how the system progressively restores demand as repairs are completed.

To further evaluate the performance and robustness of the learning crew dispatch approach, we conducted sensitivity analyses by altering the depot configurations. In scenarios 2-4 and 2-5, the locations of the original depots changed, while in scenarios 2-6 and 2-7, the number of depots was increased to seven, each with a distinct set of locations and the number of crew teams. The results, summarized in Table~\ref{table:Sensitivity}, demonstrate that the proposed method consistently identified feasible and efficient crew dispatch solutions under varying spatial configurations. In all cases, the model maintained compliance with routing constraints and achieved restoration outcomes within acceptable computation times, highlighting the adaptability and computational efficiency of the transformer-based approach.

\begin{table}[h]
\centering
\scriptsize
\setlength{\tabcolsep}{10pt}
\caption{\scriptsize Sensitivity Analysis of the Learning Crew Dispatch}
\label{table:Sensitivity}
\begin{tabular}{lcccc}
\toprule
\textbf{Scenario} & \textbf{\# Depot} & \textbf{\# Crew per Depot} & \textbf{Comp. Time} \\
\midrule
Sce. 2-4 & 5 & 1   & 13 sec \\
Sce. 2-5 & 5 & 3  & 13 sec \\
Sce. 2-6 & 7 & 1   & 13 sec \\
Sce. 2-7 & 7 & 3   & 14 sec \\
\bottomrule
\end{tabular}
\end{table}
To compare the runtime of the proposed method against other commonly used approaches on the 123-bus system, we evaluated three different earthquake intensity levels, each increasing the number of damaged components and thus the complexity of the crew dispatch problem. As the problem size grows, classical solvers like CPLEX and Gurobi often struggle to converge in a reasonable time. They may require relaxing the duality gap (e.g., up to 30\%) to return a suboptimal result. In contrast, the transformer-based method maintains fast and reliable convergence without compromising solution quality. Table~\ref{table:comparison} shows that the transformer-based approach achieves substantial reductions in computation time across all cases.

\begin{table}[h]
\centering
\scriptsize
\caption{Running Time of Crew Dispatch Solvers for 123-Bus System}
\label{table:comparison}
\renewcommand{\arraystretch}{1.1}
\begin{tabularx}{\linewidth}{>{\centering\arraybackslash}X 
                                  >{\centering\arraybackslash}X 
                                  >{\centering\arraybackslash}X 
                                  >{\centering\arraybackslash}X 
                                  >{\centering\arraybackslash}X}
\toprule
\textbf{Intensity (R)} & \shortstack{\textbf{CPLEX}\\(min:sec)} & \shortstack{\textbf{Gurobi}\\(min:sec)} & \shortstack{\textbf{GA}\\(min:sec)} & \shortstack{\textbf{Learning CD}\\(min:sec)} \\
\midrule
6.5 & 38:07 (176) & 30:27 (141) & 13:12 (61) & \cellcolor{gray!20}00:13 \\
7.5 & 138:51 (595) & 151:16 (649) & 32:35 (140) & \cellcolor{gray!20}00:14 \\
8.5 & 197:22 (846) & 231:03 (990) & 58:43 (252) & \cellcolor{gray!20}00:14 \\
\bottomrule
\end{tabularx}
\begin{tablenotes}
\scriptsize
\item Note: Values in parentheses represent the runtime factor relative to the Transformer solver.
\end{tablenotes}
\end{table}

\begin{figure}[]
    \centering
    \includegraphics[width=0.65\columnwidth]{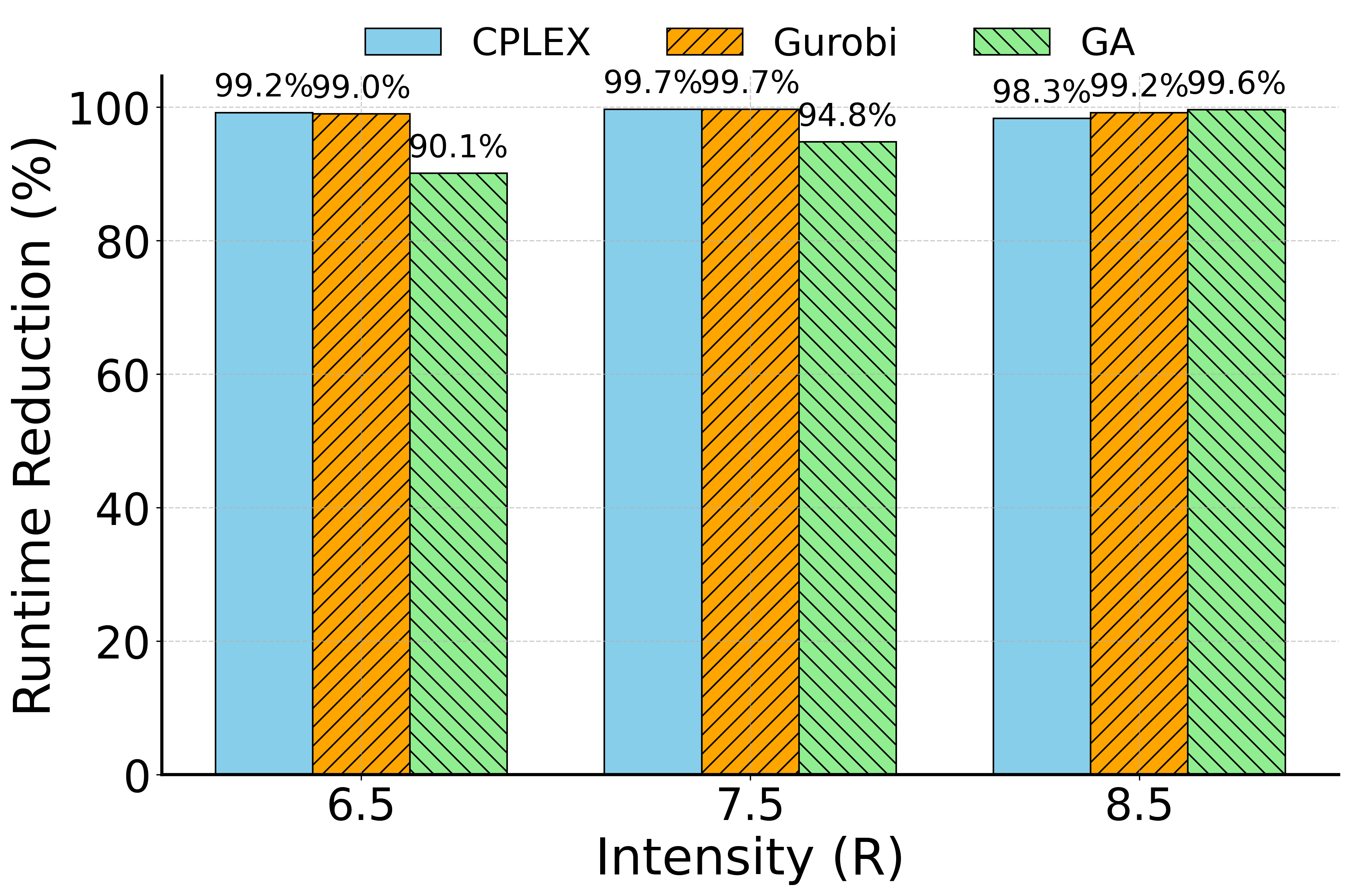} 
    \caption{\scriptsize Runtime reduction percentage across solvers for different intensity levels.}
    \label{fig:runtime_reduction}
\end{figure}

Note that while the other approaches need a long runtime, the proposed method remains efficient even as the system becomes more stressed. Fig.~\ref{fig:runtime_reduction} visualizes the percentage of runtime reduction achieved by the proposed method compared to other approaches across different intensity levels.

\begin{figure}[]
    \centering
    \includegraphics[width=0.7\columnwidth]{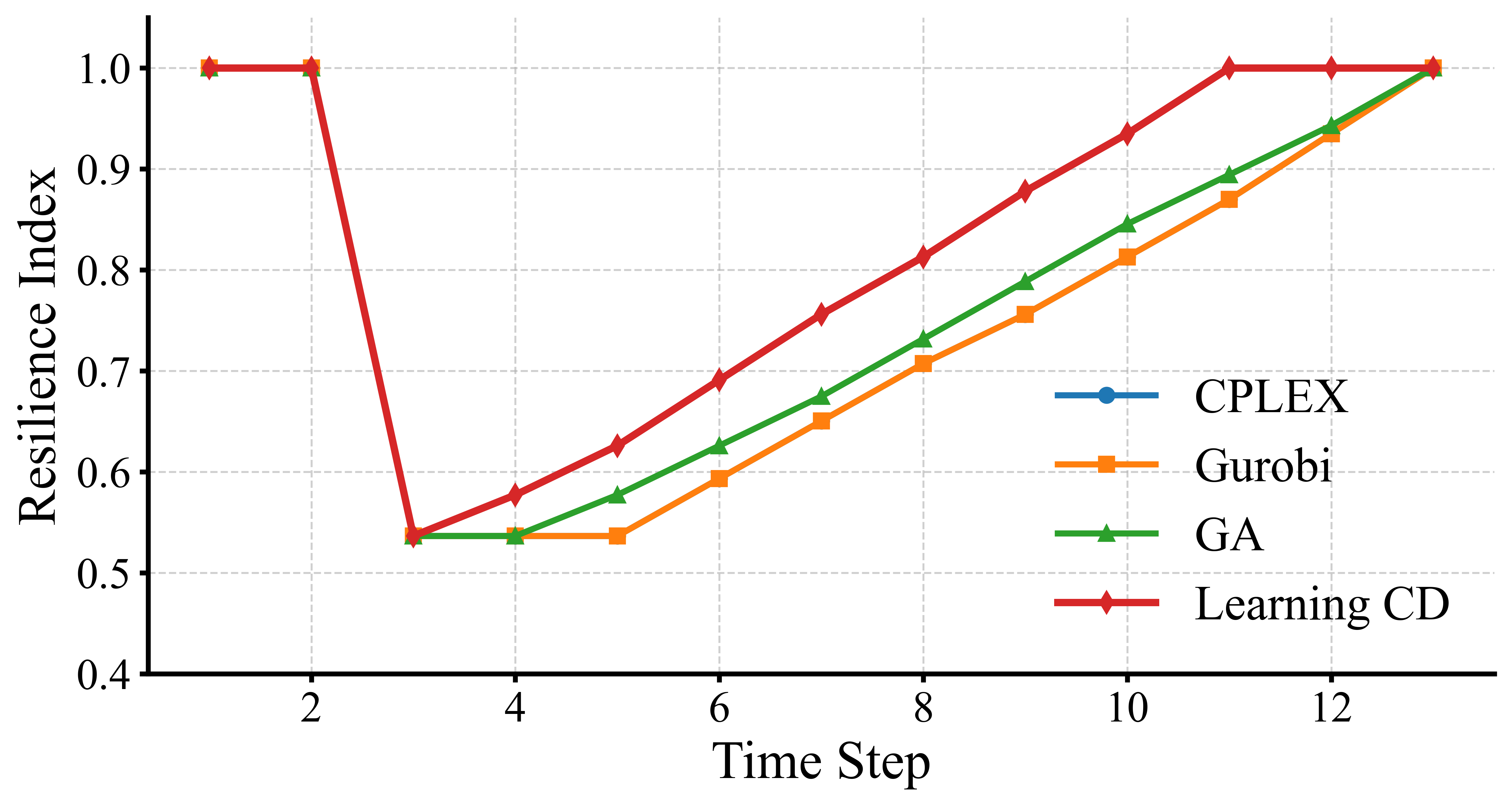} 
    \caption{\scriptsize 123-bus system resilience curves under 8.5 Richter earthquake.}
    \label{fig:123resilience}
\end{figure}

The proposed method achieves up to 99.7\% runtime reduction, marking a significant improvement over classical approaches. In post-disaster scenarios such as earthquakes or hurricanes, the heavy computational demands of traditional methods can hinder timely crew dispatch. In contrast, the proposed approach enables near-optimal decisions rapidly, allowing faster system recovery with reduced loss. Additional delays in decision-making directly translate into prolonged recovery periods, slowing down the system’s return to normal operation. For instance, in the 123-bus system, Fig.~\ref{fig:123resilience} demonstrates this effect through the resilience index, which quantifies the operational state of the network as the percentage of components functioning over time. The learning-based crew dispatch method enables faster system recovery than traditional solvers, showing higher resilience index values throughout. It achieves earlier restoration, reducing downtime and enhancing post-disaster response efficiency.

\vspace{-10pt}

\subsection{2869-bus System for Scalability Analysis}
A 2869-bus European gas and electric system is used to demonstrate real-world scale and evaluate the scalability of the proposed approach. The system is tested under three earthquake magnitudes: 6.5, 7.5, and 8.5 on the Richter scale. Earthquake magnitude determines the number of damaged components and, thus, the problem size. The number of damaged components corresponding to each case is reported in Table~\ref{table:Scalability}, which shows that a one-unit increase in magnitude can cause an exponential rise in system damage.

\begin{table}[]
\centering
\scriptsize
\caption{\scriptsize Running Time of Crew Dispatch Solvers for 2869-bus system}
\label{table:Scalability}
\begin{tabular}{c@{\hspace{10pt}}c@{\hspace{10pt}}c}
\toprule
\textbf{Intensity (R)} & \shortstack{\textbf{\# Damaged Components}} & \shortstack{\textbf{Learning CD (sec)}} \\
\midrule
6.5 & 156 & \cellcolor{gray!20}14 \\
7.5 & 325 & \cellcolor{gray!20}14 \\
8.5 & 673 & \cellcolor{gray!20}16 \\
\bottomrule
\end{tabular}
\end{table}

Table~\ref{table:Scalability} demonstrates the scalability of the learning crew dispatch approach for the 2869-bus system. As earthquake magnitude increases from 6.5 to 8.5, the number of damaged components grows rapidly from 156 to 673, reflecting a significant rise in problem size. Despite this exponential growth, the proposed transformer-based method maintains very short computation times, increasing only from 14 to 16 seconds. This result indicates that the method scales efficiently with system size and complexity, making it highly suitable for large-scale, real-world applications.

To provide a comparison between the proposed crew dispatch and conventional approaches, we implemented the classical methods with a maximum runtime limit of five hours and reported their corresponding objective function values, measured as the total economic loss due to ENS. CPLEX and Gurobi were unable to reach a solution with zero percent optimality gap within the five-hour limit; therefore, their results were recorded at termination. The GA-based method, on the other hand, completed its run after two hours in the case of 6.5 Richter and three hours for the other scenarios. During the solver runtime, we assume that no repair actions are executed in the field—that is, crews remain idle until the decision is obtained. Consequently, no load is restored during the waiting period. To capture this effect, we report the ENS-based monetary loss incurred during the waiting time (shown in parentheses) in addition to the ENS value corresponding to the recovery phase, as summarized in Table~\ref{table:obj2869}.  

\begin{table}[]
\centering
\scriptsize
\caption{Total \$M Loss of the 2869-bus System Due to ENS}
\label{table:obj2869}
\renewcommand{\arraystretch}{1.1}
\begin{tabularx}{\linewidth}{>{\centering\arraybackslash}X 
                                  >{\centering\arraybackslash}X 
                                  >{\centering\arraybackslash}X 
                                  >{\centering\arraybackslash}X 
                                  >{\centering\arraybackslash}X}
\toprule
\textbf{Intensity (R)} & \shortstack{\textbf{CPLEX}} & \shortstack{\textbf{Gurobi}} & \shortstack{\textbf{GA}} & \shortstack{\textbf{Learning CD}} \\
\midrule
6.5 & 5.4 + (2.1) & 5.4 + (2) & 4.6 + (0.8) & \cellcolor{gray!20} 4.1 + (0) \\
7.5 & 11.8 + (4.6) & 11.9 + (4.4) & 10 + (2.5) & \cellcolor{gray!20}9.2 + (0) \\
8.5 & 34.5 + (6) & 35.2 + (5.7) & 27.1 + (3.3) & \cellcolor{gray!20}22.2 + (0) \\
\bottomrule
\end{tabularx}
\\[4pt]
\begin{tablenotes}
\scriptsize
\item Note: Values in parentheses represent the additional cost incurred during solver runtime (waiting period) before crew dispatch begins.
\end{tablenotes}
\end{table}

The results in Table~\ref{table:obj2869} demonstrate that the proposed learning-based crew dispatch consistently achieves lower ENS-related losses compared to classical methods. While CPLEX and Gurobi incur significant additional costs due to long runtimes, and GA provides only moderate improvements, Learning CD eliminates waiting-time losses and delivers the most effective restoration performance across all earthquake intensities.

\begin{figure}[h]
    \centering
    \includegraphics[width=0.7\columnwidth]{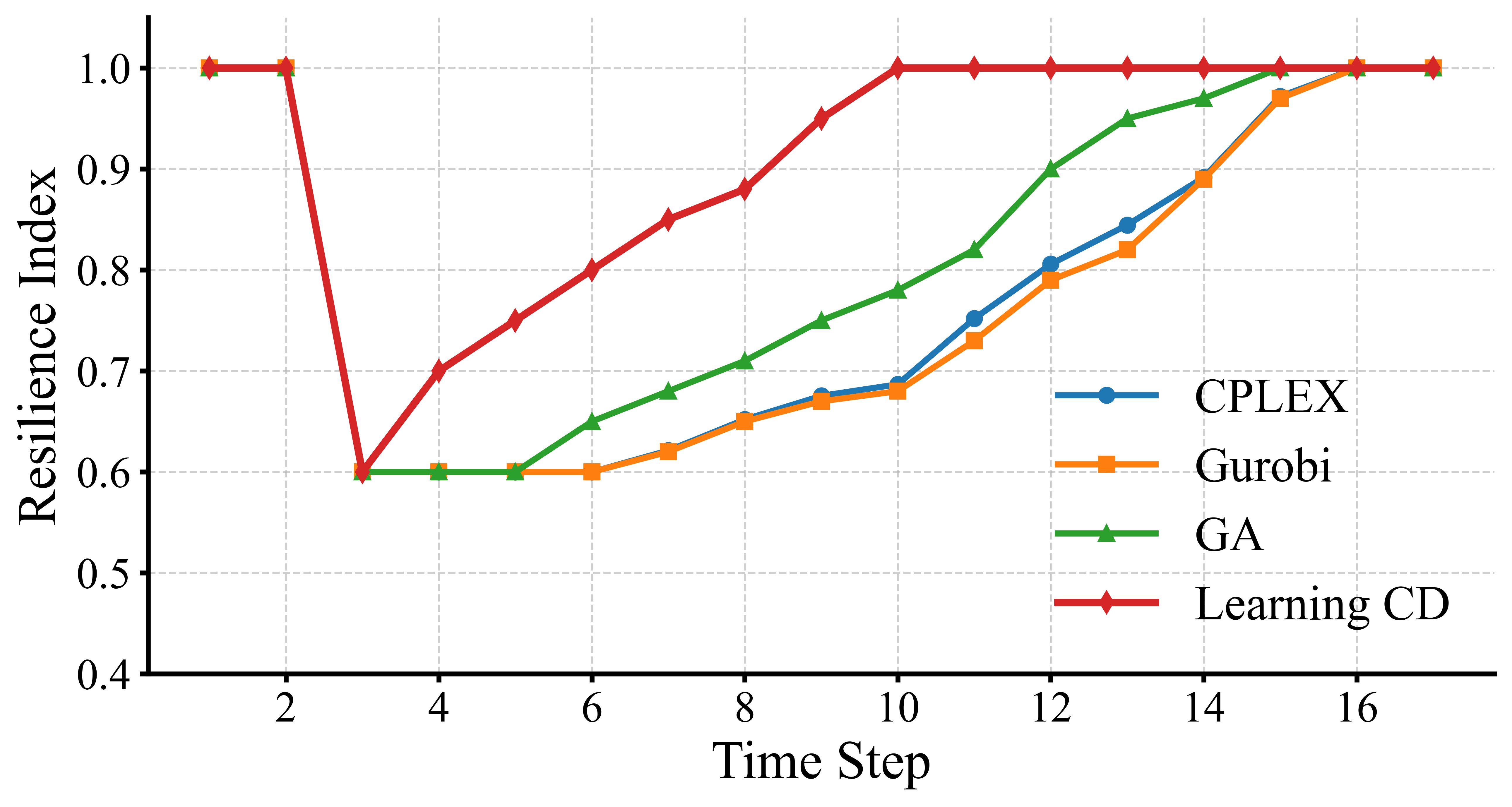} 
    \caption{\scriptsize 2869-bus system resilience curves under 8.5 Richter earthquake.}
    \label{fig:resilience}
\end{figure}

The resilience curves in Fig.~\ref{fig:resilience} demonstrate the comparative system recovery performance under the 8.5 Richter earthquake scenario. Both CPLEX and Gurobi exhibit delayed recovery trajectories due to their long computation times, resulting in slower system restoration. The GA approach achieves faster recovery than classical solvers but still lags in reaching full resilience. In contrast, the proposed learning-based crew dispatch provides a significantly steeper recovery path, restoring system performance more quickly and eliminating the additional losses associated with extended decision-making delays. These results highlight the practical advantage of the learning-based approach for real-time post-disaster restoration.

\vspace{-10pt}

\section{conclusion}
\label{sec:conclusion}
This paper presents a framework for assessing the impact of earthquakes on power distribution systems, aimed at equipping system operators with actionable insights for preparedness and response. The proposed methodology begins with realistic modeling of seismic incidents, followed by post-earthquake damage assessment and recovery planning. A region-adaptable scenario generation and reduction approach is introduced to simulate a diverse range of plausible seismic outcomes, capturing component-level vulnerabilities. To account for disrupted operating conditions—such as islanding, disconnected feeders, and out-of-service elements—a robust power flow analysis is conducted. Based on the reduced set of damage scenarios, ENS is evaluated to inform the development of an efficient and prioritized restoration strategy.

To address the computational burden associated with large-scale crew dispatch problems, we propose a learning-based dispatch model. This approach demonstrates strong scalability, efficiently solving routing problems for systems ranging from 13 to 2869 buses with negligible increase in runtime. Compared to conventional solvers such as CPLEX, Gurobi, and GAs, the proposed model achieves up to 99.7\% runtime reduction, while enhancing resilience by enabling faster and more effective recovery and restoration. These results underscore the framework’s practical value for real-time or near-real-time disaster recovery applications, offering both speed and robustness for utility-scale deployment. Future work will explore extending the framework to multi-hazard scenarios and integrating renewable energy resources into restoration planning.


\bibliographystyle{IEEEtran}
\bibliography{4th_references}

\end{document}